\documentclass[%
 amsmath,
 amssymb,
 preprint,%
]{revtex4-1}

\usepackage{graphicx}
\usepackage[caption=false]{subfig}  
\usepackage{dcolumn} 
\usepackage{bm}      

\usepackage{mathptmx} 

\usepackage{hyperref}
\hypersetup{
    colorlinks=true,
    linkcolor=blue,
    filecolor=magenta,      
    urlcolor=cyan,
    breaklinks=true
}

\DeclareMathOperator{\Imag}{Im}
\DeclareMathOperator{\Real}{Re}

\allowdisplaybreaks

\begin{document}

\pagestyle{plain}

\title{Soft Hydraulics in Channels with Thick Walls:\\ The Finite-Reynolds-Number Base State and Its Stability}

\author{Xiaojia Wang} 
 \email{wang4142@purdue.edu}
\author{Ivan C.\ Christov}
 \email[Corresponding author: ]{christov@purdue.edu} 
\affiliation{School of Mechanical Engineering, Purdue University, West Lafayette, Indiana 47907, USA}

\begin{abstract}
We analyze the linear stability of the base state of the problem of coupled flow and deformation in a long and shallow rectangular soft hydraulic conduit with a thick top wall. Specifically, the steady base state is computed at low but finite Reynolds number. Then, we show that with the upstream flux fixed and the outlet pressure set to gauge, the flow is linearly stable to infinitesimal flow-wise perturbations. Multiple oscillatory but stable eigenmodes are computed in a range of the reduced Reynolds number, $\hat{Re}$, and the so-called fluid--structure interaction (FSI) parameter, $\lambda$, indicating the stiffness of this FSI system. These results provide a framework to address, in future work, the individual effects of various aspects of two-way FSI coupling on instability and flow transition in soft hydraulic conduits.
\end{abstract}

\maketitle

\section{\label{sec:intro}Introduction}

The fluid--structure interactions (FSIs) between external or internal flows (either viscous or inviscid) and elastic structures, as well as the linear stability of such coupled mechanics problems, is a research subject with a a time-honored  history \cite{P16}. While FSI topics such as aeroelasticity \cite{BAH96} and blood flow in large arteries \cite{P80} are now quite classical, the mechanical interaction between \emph{slow} viscous flows and compliant conduits \cite{CPFY12} has opened new avenues of FSI research \cite{DS16,KCC18}, both at the microscale for, e.g., for lab-on-a-chip applications \cite{FZPN19}, and at the macroscale for, e.g., soft robotics applications  \cite{MEG17}.

In the present work, motivated by recent ``ultrafast mixing'' experimental studies in compliant microchannels \citep{VK13,KB16}, we wish to determine the linear stability of finite-Reynolds-number perturbations to the steady flow and deformation solution for FSI in a rectangular soft hydraulic conduit with a thick top wall. We derived the vanishing-Reynolds-number steady FSI solution in our previous work  \cite{WC19}. Unlike the prior study \cite{VK13}, herein we do not use experimental, computational, or other empirical information to derive our linear stability model (beyond the standard assumptions on separation of length scales, and the smallness of relevant parameters in the system). In doing so, we address the linear stability consequences of different FSI effects in soft-walled microchannels, such as the non-constant axial pressure gradient and the non-flat (deformed) base state of the flow conduit, by extending the results from our recent rigorous mathematical theory \cite{WC19}.

Furthermore, we investigate the relative importance and effect of the flow inertia (quantified by the reduced Reynolds number, $\hat{Re}$), and the compliance of the top wall (quantified by the FSI parameter, $\lambda$), on the linear stability problem. In particular, the base state is found to be stable in the range of $\hat{Re}$ and $\lambda$ considered herein, which is a typical range for microfluidic systems. We conclude with a discussion of possible extensions to the present theory.

\begin{figure}[ht]
    \centering
    \includegraphics[width=0.6\textwidth]{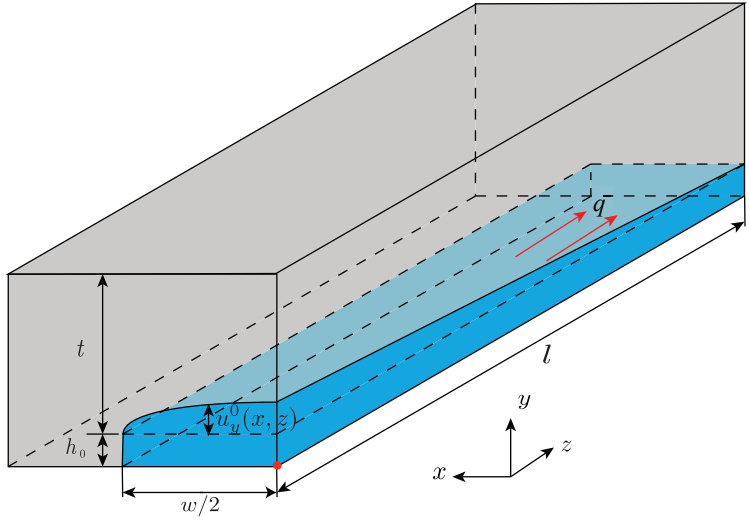}
    \caption{Diagram of one-half of an $x$-symmetric thick-walled microchannel, labelled with the dimensional variables (lower case) of the problem. The origin of the coordinate system (labeled with a red a dot) is set at the centerline ($x=0$) of the rigid bottom wall of the channel. Here, $h_0$, $w$, and $\ell$ represent the undeformed channel height, width and length, respectively, while $t$ is the top wall's thickness. The deformed fluid--solid interface is defined as $y=h_0+u_y^0(x,z)$, where the compliant top wall's $y$-displacement evaluated at $y=h_0$ is denoted by $u_y^0$. The Newtonian fluid flow, with a given volumetric flow rate $q$, is in the positive $z$-direction, as indicated by arrows, from the inlet at $z=0$ to the outlet at $z=\ell$. The reduced Reynolds number introduced in Eq.~\eqref{NS-O1} can be defined using the dimensional variables in the figure as $\hat{Re}=\epsilon Re={qh_0}/(\nu w\ell)$, where $\nu$ is the kinematic viscosity of the fluid, and $\epsilon=h_0/\ell$ is the axial aspect ratio. Reproduced and adapted with permission from Ref.~\cite{WC19} \textcopyright\ 2019 The Author(s) (X.W.\ and I.C.C.) Published by the Royal Society.}
    \label{fig:schematic}
\end{figure}

\section{\label{sec:gov_eq}Governing Equations}

To consider finite-Reynolds-number perturbations to the steady $Re=0$ base flow, we allow a finite \emph{reduced} Reynolds number: $\hat{Re} = \epsilon Re = \mathcal{O}(1)$ as $\epsilon\to0$, where $\epsilon\ll1$ is the undeformed-height-to-length ratio of the long and shallow microchannel (see Fig.~\ref{fig:schematic} for notation and schematic of the physical setup). Then, the leading-order (in $\epsilon$) governing incompressible Navier--Stokes flow equations are as follows (see Ref.~\cite{WC19} for the derivation and discussion):
\begin{subequations}\label{NS-O1}
\begin{align}
    \frac{\partial V_X}{\partial X}+\frac{\partial V_Y}{\partial Y}+\frac{\partial V_Z}{\partial Z}&=0,\label{COM-O1}\displaybreak[3]\\
    -\frac{\partial P}{\partial X}&=0,\label{COLM-X-O1}\\
    -\frac{\partial P}{\partial Y}&=0,\label{COLM-Y-O1}\\
    \hat{Re}\left(\frac{\partial V_Z}{\partial T}+V_X\frac{\partial V_Z}{\partial X}+V_Y\frac{\partial V_Z}{\partial Y}+V_Z\frac{\partial V_Z}{\partial Z}\right)&=-\frac{\partial P}{\partial Z}+\frac{\partial^2 V_Z}{\partial Y^2}\label{COLM-Z-O1}.
\end{align}
\end{subequations}
These equations, and all capital letters used herein are dimensionless. The non-dimensionalization is standard and discussed in Ref.~\cite{WC19}. For the present purposes, since we will not use the dimensional variables at all in the discussion below, we do not go over the non-dimensionalization. 
Equation~\eqref{COM-O1} is the continuity (conservation of mass) equation, which is balanced at the leading order. Equations~\eqref{COLM-X-O1}, \eqref{COLM-Y-O1}, and  \eqref{COLM-Z-O1} are the conservation of linear momentum equations in the $X$, $Y$, and $Z$ directions respectively. Owing to the long and shallow nature of the microchannel, the $X$ and $Y$ equations simply state there is no pressure gradients in those directions at the leading order in $\epsilon$, and the flow is primarily unidirectional in the $Z$ direction.

We are interested in the flow regime in which the characteristic time scale set by the compliant wall's inertia is much smaller than the characteristic flow time scale. In other words, we assume that the inertia of the elastic solid is negligible, and the unsteadiness in this FSI system is fully determined by the fluid flow. This assumption is often invoked when studying the relaxation time \cite{PYDHD09} or the start-up time \cite{MCSPS19} of compliant microchannels. Note, however, it is also possible that, in some regimes, the inertia of the compliant wall may play a role in the unsteady inflation or relaxation of the soft wall, due to the interplay between the deformation and flow  \cite{IWC20,MCSPS19}.

Here, having restricted to a prototypical microsystem in which we can neglect the inertia of the elastic wall, the displacement field developed in Ref.~\cite{WC19} can be transferred smoothly into the unsteady problem. Specifically, for a thick top wall, as considered herein, with $(t/w)^2\gg1$, the (dimensionless) deformation profile at the fluid--solid interface (again, see Ref.~\cite{WC19} for the derivation and discussion) is 
\begin{equation}\label{uyinf_dimless}
    U_Y^0(X,Z,T)=P(Z,T)\underbrace{\sum_{m=1}^{\infty}\frac{2A_m}{m\pi}\sin\left[m\pi\left(X+\frac{1}{2}\right)\right]}_{=:\mathfrak{G}(X)},
\end{equation}
where $A_m=\frac{2}{m\pi}[1-(-1)^m]$. Thus, the deformed channel height is 
\begin{equation}\label{H}
    H(X,Z,T) = 1 + \lambda U_Y^0(X,Z,T) \stackrel{\text{by Eq.~\eqref{uyinf_dimless}}}{=} 1 + \lambda P(Z,T)\mathfrak{G}(X).
\end{equation}
Here, $\lambda = u_c/h_0$, which is the ratio of the characteristic deformation scale $u_c$ of the elastic solid to the undeformed channel height $h_0$, is termed the \emph{FSI parameter}; for $\lambda=0$, there is no deformation, while for $\lambda=\mathcal{O}(1)$ significant FSI-induced deformation of the flow conduit occurs.

Unlike the case in Ref.~\cite{WC19}, here we retain the $\hat{Re}$ terms as $\epsilon\to0$, which yields a nonlinear governing equation~\eqref{COLM-Z-O1} for $V_Z$. To make progress, it is standard to integrate Eqs.~\eqref{NS-O1} across a deformed axial cross-section (fixed $Z$) and to introduce the flow rate $$Q(Z,T) \equiv \int_{-1/2}^{+1/2}\int_{0}^{H(X,Z,T)} V_Z(X,Y,Z,T) \,dY\,dX$$ into the formulation (see, e.g., \cite{SWJ09} and the references therein). However, after this integration, we still need a relation between $V_Z$ and $Q$ to deal with the integral in $Y$. Here, motivated by prior studies on inertial fluid effects in microchannels \cite{SWJ09,IWC20}, we apply the von K\'arm\'an--Polhausen approximation \cite{panton} for the velocity profile:
\begin{equation}\label{Vz-Q}
    V_Z(X,Y,Z,T) = \frac{6Q\big[H(X,Z,T)-Y\big]Y}{\int_{-1/2}^{+1/2}H(X,Z,T)^3 \, dX}.
\end{equation}
Essentially, this assumption enforces a parabolic (Poiseuille) profile in each axial cross-section, while simultaneously accounting for the flow-wise variation of the height $H$. Also, note that the assumed closure relation~\eqref{Vz-Q} is consistent with the previous result \cite{WC19} in the limit $\hat{Re}\to 0$. Furthermore, the kinematic boundary condition is imposed at the moving fluid--solid interface:
\begin{equation}\label{kinematic}
    \frac{\partial H}{\partial T}=\left.V_Y\right|_{Y=H(X,Z,T)}.
\end{equation}

Then, performing the cross-sectional integration of the governing equations~\eqref{NS-O1}, substituting the ansatz~\eqref{Vz-Q}, using the condition~\eqref{kinematic}, and simplifying, we obtain 
\begin{subequations}\label{inteq}\begin{align}
    \frac{\partial Q}{\partial Z}+\lambda\mathfrak{I}_1\frac{\partial P}{\partial T}&=0,\label{inteq1}\\
    \hat{Re}\left[\frac{\partial Q}{\partial T}+\frac{6}{5}\frac{\partial}{\partial Z}\left(\frac{\mathfrak{C}}{\mathfrak{B}^2}Q^2\right)\right]&=-\frac{\partial P}{\partial Z}(1+\lambda\mathfrak{I}_1 P)- \frac{12\mathfrak{A}}{\mathfrak{B}}Q\label{inteq2},
\end{align}\end{subequations}
where
\begin{subequations}
\begin{align}
    \mathfrak{I}_i=&\int_{-1/2}^{+1/2}\mathfrak{G}^i(X)\,dX, \quad i = 1,2,\hdots,5,\label{Ii}\\
    \mathfrak{A}[P(Z)]=&1+\lambda \mathfrak{I}_1 P(Z),\label{A}\\
    \mathfrak{B}[P(Z)]=&1+3\lambda\mathfrak{I}_1P(Z)+3\lambda^2\mathfrak{I}_2P^2(Z)+\lambda^3\mathfrak{I}_3 P^3(Z)\label{B}\\
    \mathfrak{C}[P(Z)]=&1+5\lambda\mathfrak{I}_1P(Z)+10\lambda^2\mathfrak{I}_2P^2(Z)+10\lambda^3\mathfrak{I}_3 P^3(Z)+5\lambda^4\mathfrak{I}_4P^4(Z)+\lambda^5\mathfrak{I}_5P^5(Z).\label{C}
\end{align}
\end{subequations}
Equations \eqref{inteq} and \eqref{H} describe the coupling between the fluid flow and the elastic wall's deformation. Note that $H(X,Z,T)$ is completely determined by the pressure profile, $P(Z,T)$, because $\mathfrak{G}(X)$ is a known function defined by the Fourier series in Eq.~\eqref{uyinf_dimless}. Likewise, the constants $\{\mathfrak{I}_i\}_{i=1}^5$ are known; their values are pre-computed and listed in Table~\ref{tab:table-I}. Meanwhile, $\mathfrak{A}$, $\mathfrak{B}$ and $\mathfrak{C}$ are functionals of the pressure $P$ and, thus, implicitly functions of $Z$.

\begin{table}
    \caption{\label{tab:table-I} The values of the constants $\{\mathfrak{I}_i\}_{i=1}^5$ defined by Eq.~\eqref{Ii}.}
    \begin{ruledtabular}
    \begin{tabular}{c c c c c}
    $\mathfrak{I}_1$ & $\mathfrak{I}_2$ & $\mathfrak{I}_3$ & $\mathfrak{I}_4$ & $\mathfrak{I}_5$ \\
    \hline
    0.542710 & 0.333333 & 0.215834 & 0.143959 & 0.097864\\
    \end{tabular}
    \end{ruledtabular}
\end{table}

Fixing the flow rate upstream, and keeping the outlet of the channel open to atmospheric conditions, we can impose the following boundary conditions:
\begin{equation}\label{bc}
    Q|_{Z=0}=1, \qquad P|_{Z=1}=0.
\end{equation}
Note that no restrictions are imposed on the wall's deformation at the inlet and outlet. Those would require a matched asymptotic calculation taking into account axial bending (see, e.g., Ref.~\cite{AC18b} for a discussion of this issue in the context of a slender microtube), which is beyond the scope of the present work.

\subsection{The Base State at Finite $\hat{Re}$}
At steady state, the boundary conditions \eqref{bc} on the flow rate indicates that $Q(Z)\equiv1$, while $P_0(Z)$ and $H_0(X,Z)$ should satisfy
\begin{subequations}\label{steady}
\begin{align}
    \frac{d}{d Z}\left[ \hat{Re}\frac{6}{5}\frac{\mathfrak{C}}{\mathfrak{B}^2} + \left(1+\frac{1}{2}\lambda\mathfrak{I}_1 P_0\right)P_0 \right]=&-\frac{12\mathfrak{A}}{\mathfrak{B}}\label{steady1},\\
    H_0(X,Z)=&1+\lambda P_0(Z)\mathfrak{G}(X).\label{steady2}
\end{align}   
\end{subequations}
The unknown in Eqs.~\eqref{steady} is $P_0(Z)$, subject to the outlet boundary condition
\begin{equation}\label{steadybc}
P_0(Z=1)=0.
\end{equation}
If $\hat{Re}\to 0$, Eq.~\eqref{steady1} can be rewritten as  $-(\mathfrak{B}/12){d P_0}/{d Z}  =1$, where $\mathfrak{B}[P(Z)]$ is given by Eq.~\eqref{B}. This ordinary differential equation can be easily shown to match the previous result in Ref.~\cite{WC19}.

Equation \eqref{steady1} subject to Eq.~\eqref{steadybc} are solved together numerically as a ``final value problem'' using the classical fourth-order Runge--Kutta (RK4) method implemented using the python package SciPy \cite{SciPy}. In particularly, within each step of the RK4 method, a nonlinear algebraic problem must be solved because the  functionals $\mathfrak{A}$, $\mathfrak{B}$ and $\mathfrak{C}$ depend on the solution $P_0(Z)$. This nonlinear solution step is accomplished using \texttt{optimize.fsolve} from SciPy. The scheme is validated for $\hat{Re}=0$ against the analytical result from Ref.~\cite{WC19}. 

\begin{figure}[b]
    \centering
    \subfloat[]{\includegraphics[width=0.49\textwidth]{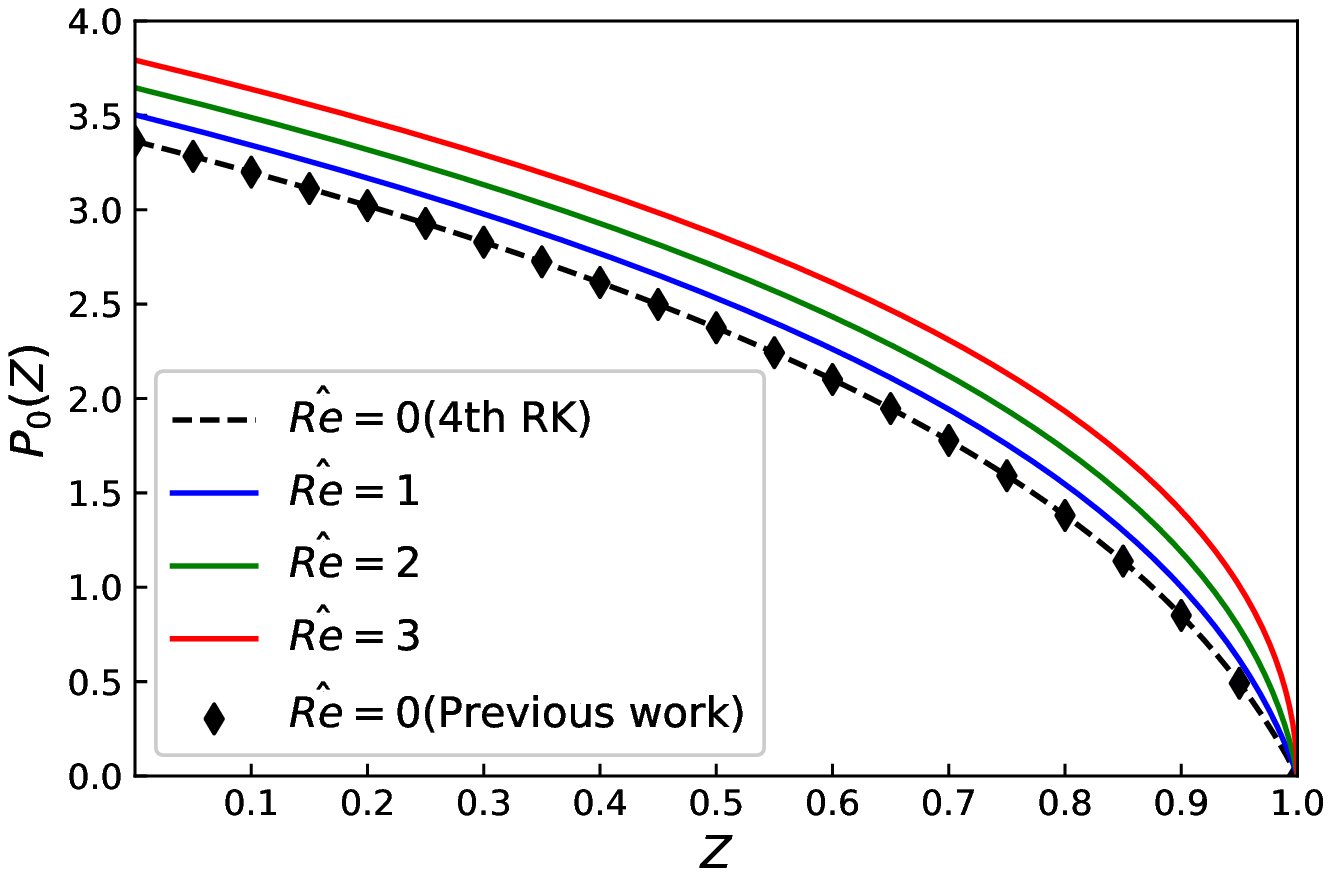}}
    \subfloat[]{\includegraphics[width=0.49\textwidth]{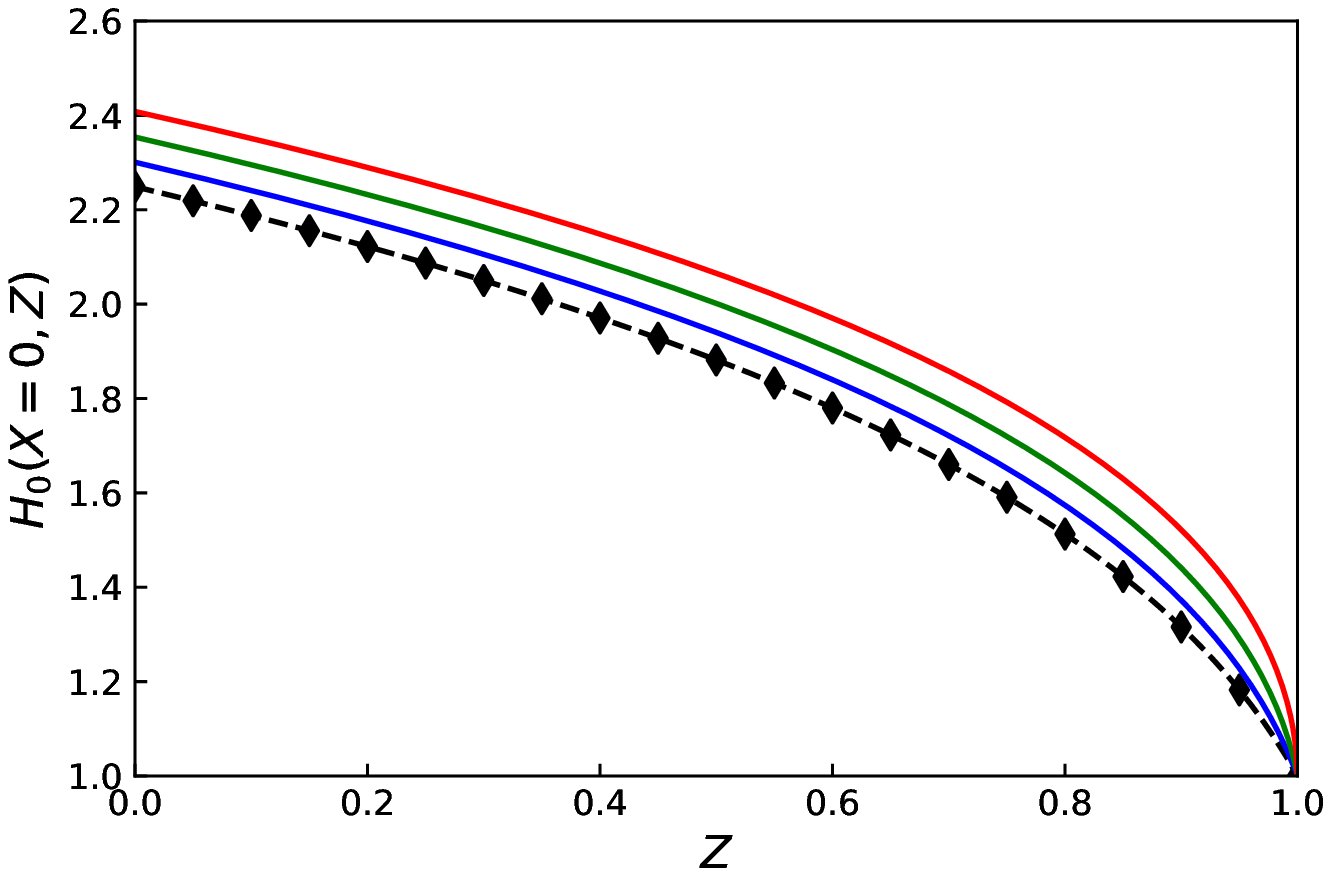}}
    \caption{Finite $\hat{Re}$ base state. (a) The steady pressure distribution $P_0$ along the flow-wise $Z$ direction. (b) The steady deformed channel shape $H_0$ along the mid-plane, $X=0$. Both panels show a set of different $\hat{Re}$ with $\lambda=0.5$ fixed. The pressure distribution for the case $\hat{Re}=0$,  computed with the present RK4 numerical method (dashed curve),  agrees exactly with the analytical result from Ref.~\cite{WC19} (symbols).}
    \label{fig:P0H0}
\end{figure}

As shown in Fig.~\ref{fig:P0H0}(a), we observe that the inclusion of flow inertia ($\hat{Re}=\mathcal{O}(1)$) results in a larger total pressure drop, $\Delta P \equiv P(1)-P(0)$, and a steeper pressure gradient $dP/dZ$ at the outlet ($Z=1$). After obtaining the pressure distribution $P_0(Z)$, the shape of the deformed channel $H_0(X,Z)$ is just a linear function of $P_0(Z)$ found from Eq.~\eqref{steady2}. Thus, as shown in Fig.~\ref{fig:P0H0}(b), the interface deformation at the channel mid-plane, $X=0$, has an identical shape to the pressure distribution.

\section{Perturbation and Linear Stability Problem}
Let us introduce the following perturbations to the steady finite-$\hat{Re}$ base state $\{Q=1,P=P_0(Z)\}$ derived in the previous section:
\begin{subequations}\label{perturb}
\begin{align}
    Q(Z,T)=&1+\alpha \widetilde{Q}(Z,T),\\
    P(Z,T)= &P_0(Z)+\alpha \widetilde{P}(Z,T),
\end{align}
\end{subequations}
where $\alpha\ll 1$ is an arbitrary small parameter quantifying the magnitude of the axial perturbations $\widetilde{Q}$ and $\widetilde{P}$. Then, it follows that 
\begin{equation}
    H(X,Z,T) = H_0(X,Z) + \alpha \lambda \widetilde{P}(Z,T)\mathfrak{G}(X).
\end{equation}
Since the actual boundary conditions were imposed on the base state, the perturbations should satisfy homogeneous boundary conditions:
\begin{equation}\label{perturb_bc1}
    \widetilde{Q}|_{Z=0}=0, \qquad \widetilde{P}|_{Z=1}=0.
\end{equation}

Next substituting Eqs.~\eqref{perturb} into the governing equations~\eqref{inteq}, and only keeping terms up to $\mathcal{O}(\alpha)$, we obtain the linearized equations governing the evolution of perturbations:
\begin{subequations}\label{stability}
\begin{align}
    \frac{\partial \widetilde{Q}}{\partial Z}+\lambda\mathfrak{I}_1\frac{\partial \widetilde{P}}{\partial T}&=0,\label{stability1}\\
    \hat{Re}\frac{\partial\widetilde{Q}}{\partial T} +\frac{6}{5}\hat{Re} \frac{\partial}{\partial Z}\left[\frac{2\mathfrak{C}_0}{\mathfrak{B}_0^2}\widetilde{Q}+\left(\frac{\mathfrak{C}_0^{\prime}}{\mathfrak{B}_0^2}
    -\frac{2\mathfrak{B}_0^{\prime}\mathfrak{C}_0}{\mathfrak{B}_0^3}\right)\widetilde{P}\right]\label{stability2}
    &= -12\frac{\mathfrak{A}_0}{\mathfrak{B}_0}\widetilde{Q}\\
    &\phantom{=}+\left(-\lambda\mathfrak{I}_1\frac{dP_0}{dZ}-12\frac{\lambda\mathfrak{I}_1}{\mathfrak{B}_0}+12\frac{\mathfrak{A}_0\mathfrak{B}_0^{\prime}}{\mathfrak{B}_0^2}\right)\widetilde{P}-\mathfrak{A}_0\frac{\partial\widetilde{P}}{\partial Z},\nonumber
\end{align}
\end{subequations}
where $\mathfrak{A}_0 \equiv \mathfrak{A}[P_0(Z)]$,  $\mathfrak{B}_0\equiv \mathfrak{B}[P_0(Z)]$ and  $\mathfrak{C}_0\equiv \mathfrak{C}[P_0(Z)]$ are evaluated via Eqs.~\eqref{A}--\eqref{C}, and
\begin{subequations}
\begin{align}
    \mathfrak{B}_0^{\prime}=&3\lambda\mathfrak{I}_1+6\lambda^2\mathfrak{I}_2P_0(Z)+3\lambda^3\mathfrak{I}_3P_0^2(Z),\label{B0prime}\\
    \mathfrak{C}_0^{\prime}=&5\lambda\mathfrak{I}_1+20\lambda^2\mathfrak{I}_2P_0(Z)+30\lambda^3\mathfrak{I}_3P_0^2(Z)+20\lambda^4\mathfrak{I}_4P_0^3(Z)+5\lambda^5\mathfrak{I}_5P_0^4(Z).\label{C0prime}
\end{align}
\end{subequations}
Note that the variables with the subscripts ``0" are obtained from the base state solution discussed in the previous section. Thus, they are known for the purposes of the upcoming linear stability calculation.

We restrict our analysis to asymptotic stability of modal perturbations (excluding any effects of transient growth arising from fact that the base state is non-constant and the linear operator is non-normal \cite{Sch07}). To this end, let
\begin{equation}\label{perturb2}
    \widetilde{Q}(Z,T)=Q_1(Z)\mathrm{e}^{-\mathrm{i}\omega T}, \qquad \widetilde{P}(Z,T)=P_1(Z) \mathrm{e}^{-\mathrm{i}\omega T}.
\end{equation}
Further applying $dQ_1/dZ=\mathrm{i}\omega\lambda\mathfrak{I}_1P_1$, Eqs.~\eqref{stability} can be rewritten in the matrix form:
\begin{equation}\label{stab_matrix}
    \begin{pmatrix}
    \frac{d}{dZ} & 0\\[1mm] \mathcal{L}_Q & \mathcal{L}_P
    \end{pmatrix}\begin{pmatrix} Q_1 \\[1mm] P_1\end{pmatrix}
    =\omega\begin{pmatrix}0 & \mathrm{i}\lambda\mathfrak{I}_1\\[1mm] \mathrm{i}\hat{Re} &  -\mathrm{i}\frac{6}{5}\hat{Re}\frac{2\mathfrak{C}_0}{\mathfrak{B}_0^2}\lambda\mathfrak{I}_1
\end{pmatrix}\begin{pmatrix} Q_1 \\[1mm] P_1\end{pmatrix},
\end{equation}
where we have defined the following operators for convenience:
\begin{subequations}
\begin{align}
    \mathcal{L}_Q =& \frac{6}{5}\hat{Re} \frac{d}{dZ}\left(\frac{2\mathfrak{C}_0}{\mathfrak{B}_0^2}\right)+12\frac{\mathfrak{A}_0}{\mathfrak{B}_0},\\[1mm]
    \mathcal{L}_P =& \frac{6}{5}\hat{Re}\left[\frac{d}{dZ}\left(\frac{\mathfrak{C}_0^{\prime}}{\mathfrak{B}_0^2}-\frac{2\mathfrak{C}_0\mathfrak{B}_0^{\prime}}{\mathfrak{B}_0^3}\right)+\left(\frac{\mathfrak{C}_0^{\prime}}{\mathfrak{B}_0^2}-\frac{2\mathfrak{C}_0\mathfrak{B}_0^{\prime}}{\mathfrak{B}_0^3}\right)\frac{d}{dZ}\right]+\lambda\mathfrak{I}_1\frac{d P_0}{dZ}+12\frac{\lambda\mathfrak{I}_1}{\mathfrak{B}_0}-12\frac{\mathfrak{A}_0\mathfrak{B}_0^{\prime}}{\mathfrak{B}_0^2}+\mathfrak{A}_0\frac{d}{dZ}.
\end{align}
\end{subequations}
The corresponding boundary conditions, obtained from Eq.~\eqref{perturb_bc1}, are
\begin{equation}\label{stab_bc1}
    Q_1(0)=0, \qquad P_1(1)=0.
\end{equation}
Substituting the latter into Eqs.~\eqref{stab_matrix}, we obtain two further boundary conditions:
\begin{equation}\label{stab_bc2}
    \left.\frac{d Q_1}{dZ}\right|_{Z=1}=0, \qquad \left.\left[\mathcal{L}_P P_1+\frac{6}{5}\hat{Re}\frac{2\mathfrak{C}_0}{\mathfrak{B}_0^2}\frac{dQ_1}{dZ}\right]\right|_{Z=0}=0.
\end{equation}

Equation~\eqref{stab_matrix} and the boundary conditions in Eqs.~\eqref{stab_bc1} and \eqref{stab_bc2} constitute a \emph{generalized eigenvalue problem}, in which $\omega\in\mathbb{C}$ is the eigenvalue and $[Q_1,P_1]^\top$ is the eigenfunction. The system is said to be linearly unstable if there exist eigenvalues with $\Imag(\omega)>0$ for a combination of the parameters. To solve this eigenvalue problem, we shall employ the Chebyshev pseudospectral numerical method. In this way, we can resolve the eigenspectra to determine if the system exhibits linear stability (or instability).

\section{Results and Discussion}

The Chebyshev pseudospectral method \cite{SH01,Boyd00} for the linear stability problem is implemented as described in  \cite{IWC20}, using the python package SciPy \cite{SciPy}. Simply speaking, the eigenfunctions $Q_1$ and $P_1$ are approximated with an $N$-th degree polynomial each, then the generalized eigenvalue problem (Eqs.~\eqref{stab_matrix}, \eqref{stab_bc1} and \eqref{stab_bc2}) is discretized by enforcing the satisfaction of the equations at $N+1$ Gauss--Lobatto points. Specifically, Eq.~\eqref{stab_matrix} is required to be satisfied at $N-1$ interior Gauss--Lobatto points while the boundary conditions \eqref{stab_bc1} and \eqref{stab_bc2} are imposed at the two end points. Furthermore, since the boundary conditions are homogeneous, the right-hand-side matrix in Eq.~\eqref{stab_matrix} is singular.

\begin{figure}
    \centering
    \subfloat[$\hat{Re}=0.01$]{\includegraphics[width=0.49\textwidth]{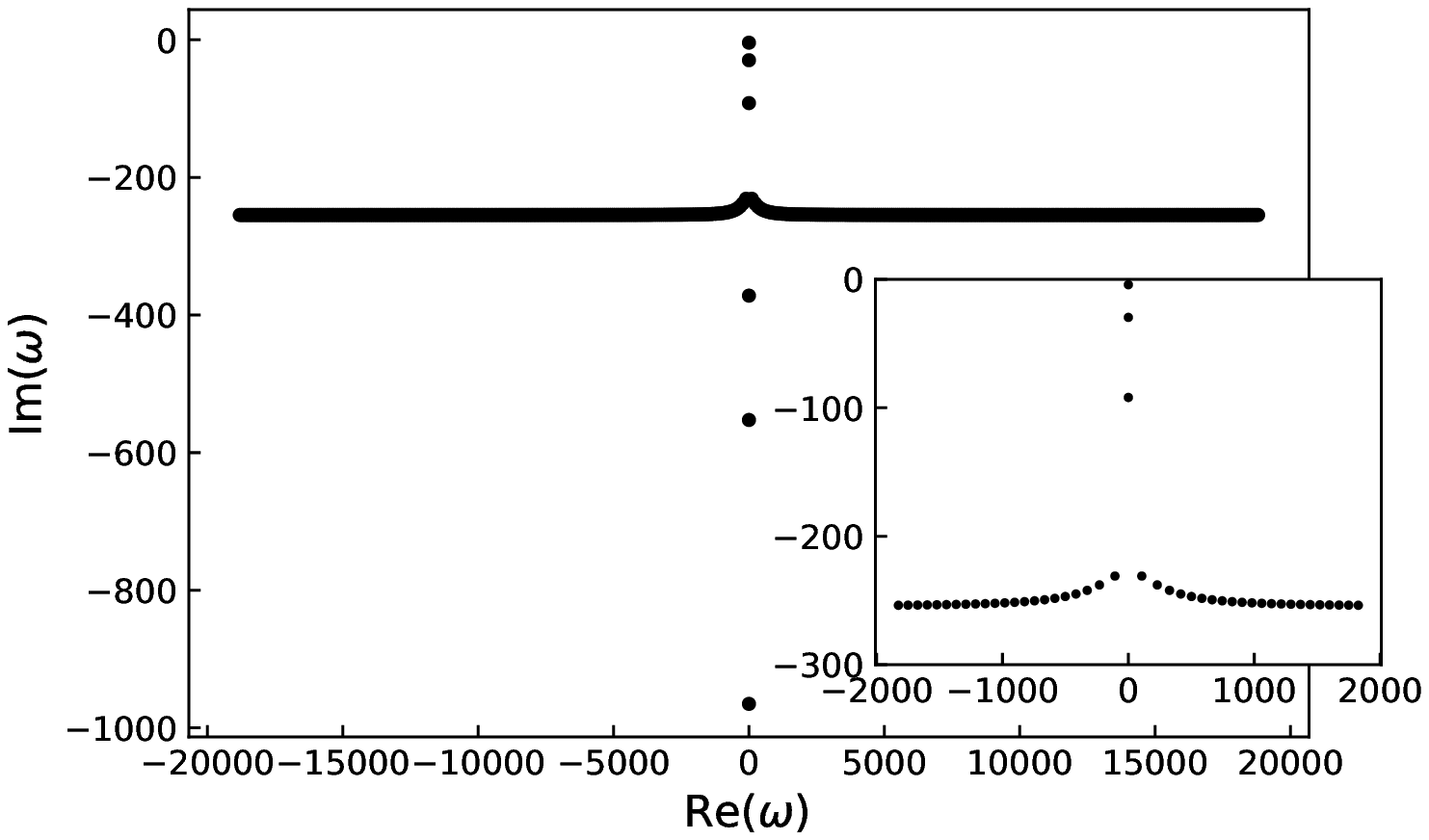}}
    \subfloat[$\hat{Re}=0.1$]{\includegraphics[width=0.49\textwidth]{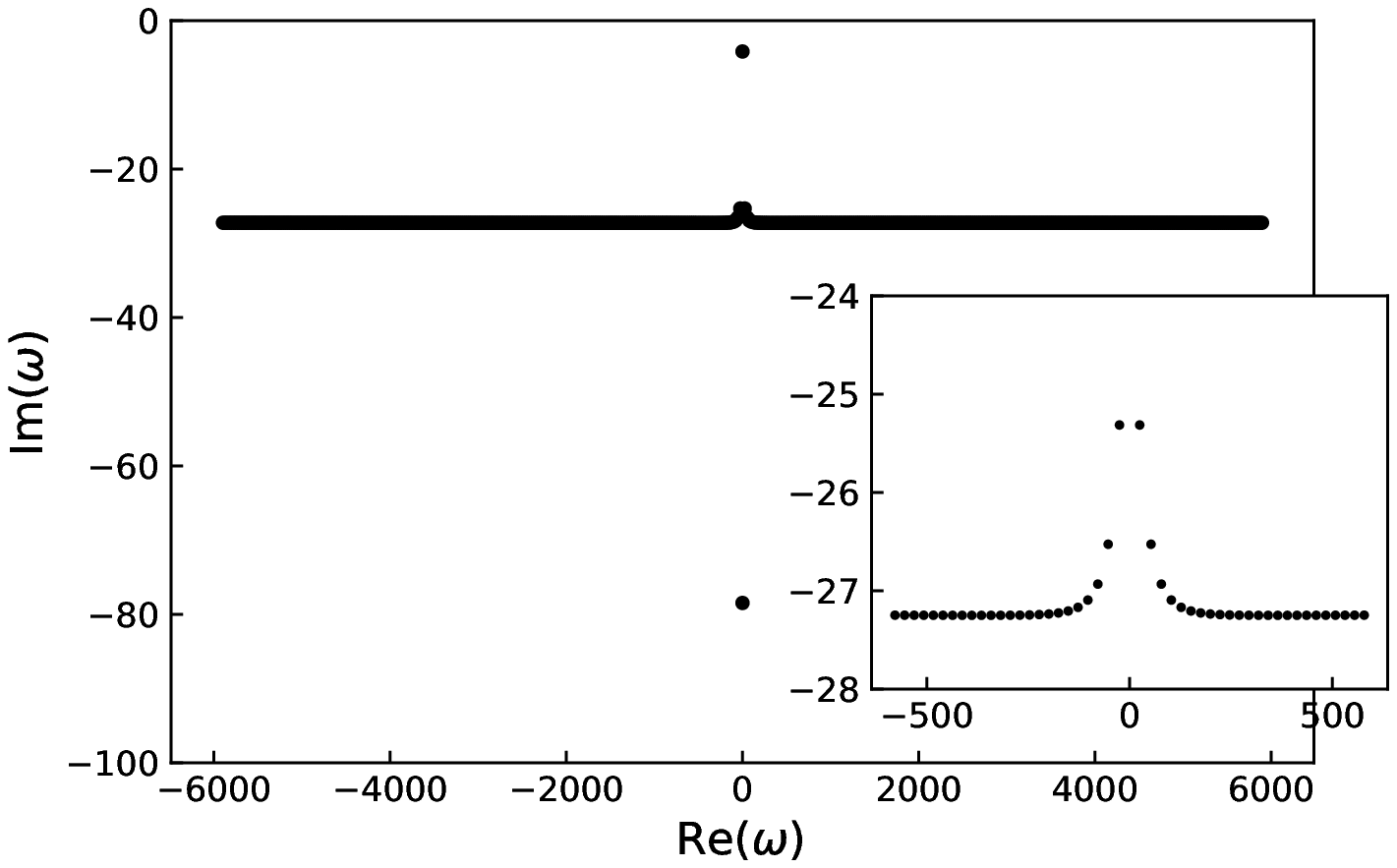}}\\
    \subfloat[$\hat{Re}=1$]{\includegraphics[width=0.49\textwidth]{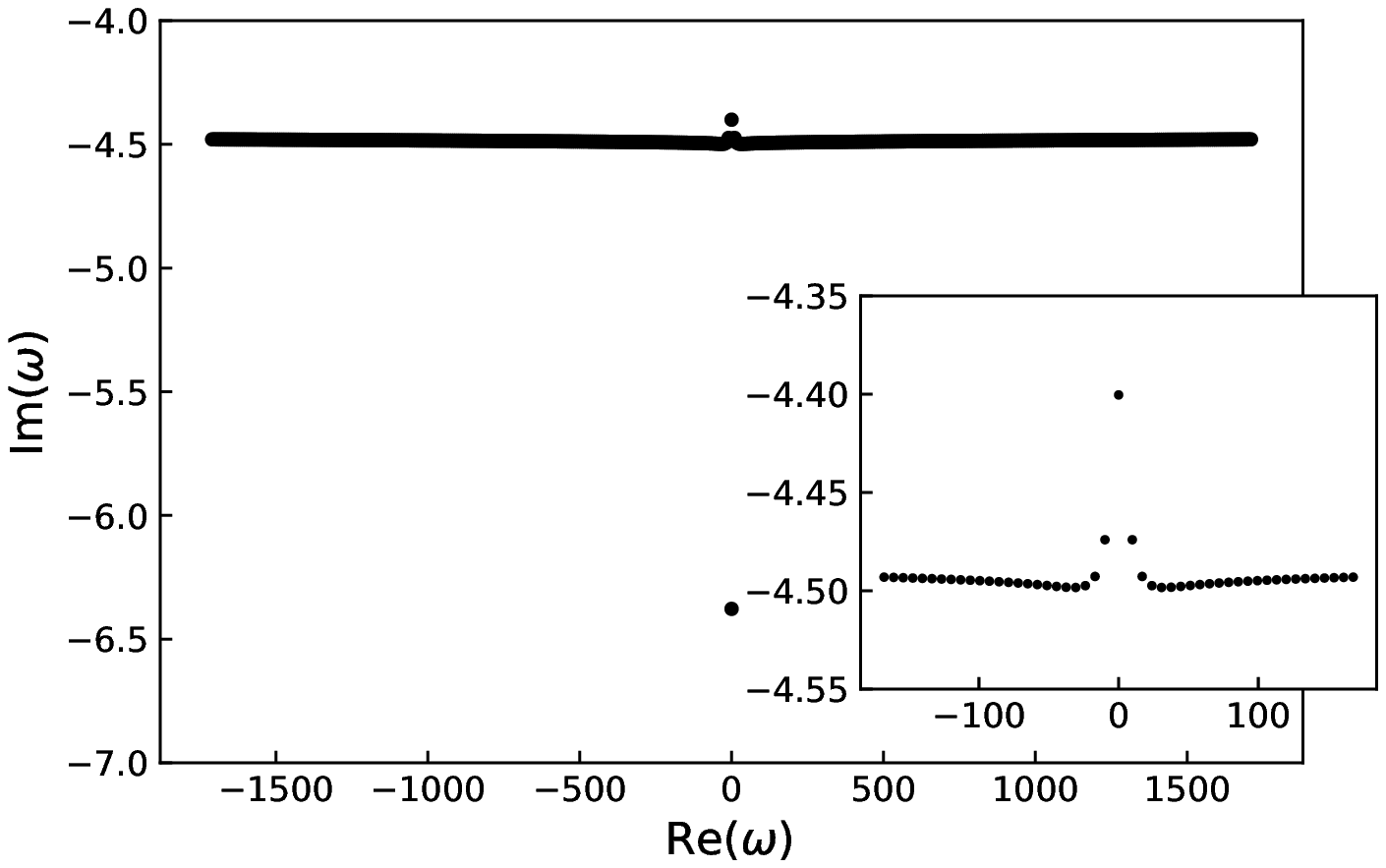}}
    \subfloat[$\hat{Re}=3$]{\includegraphics[width=0.49\textwidth]{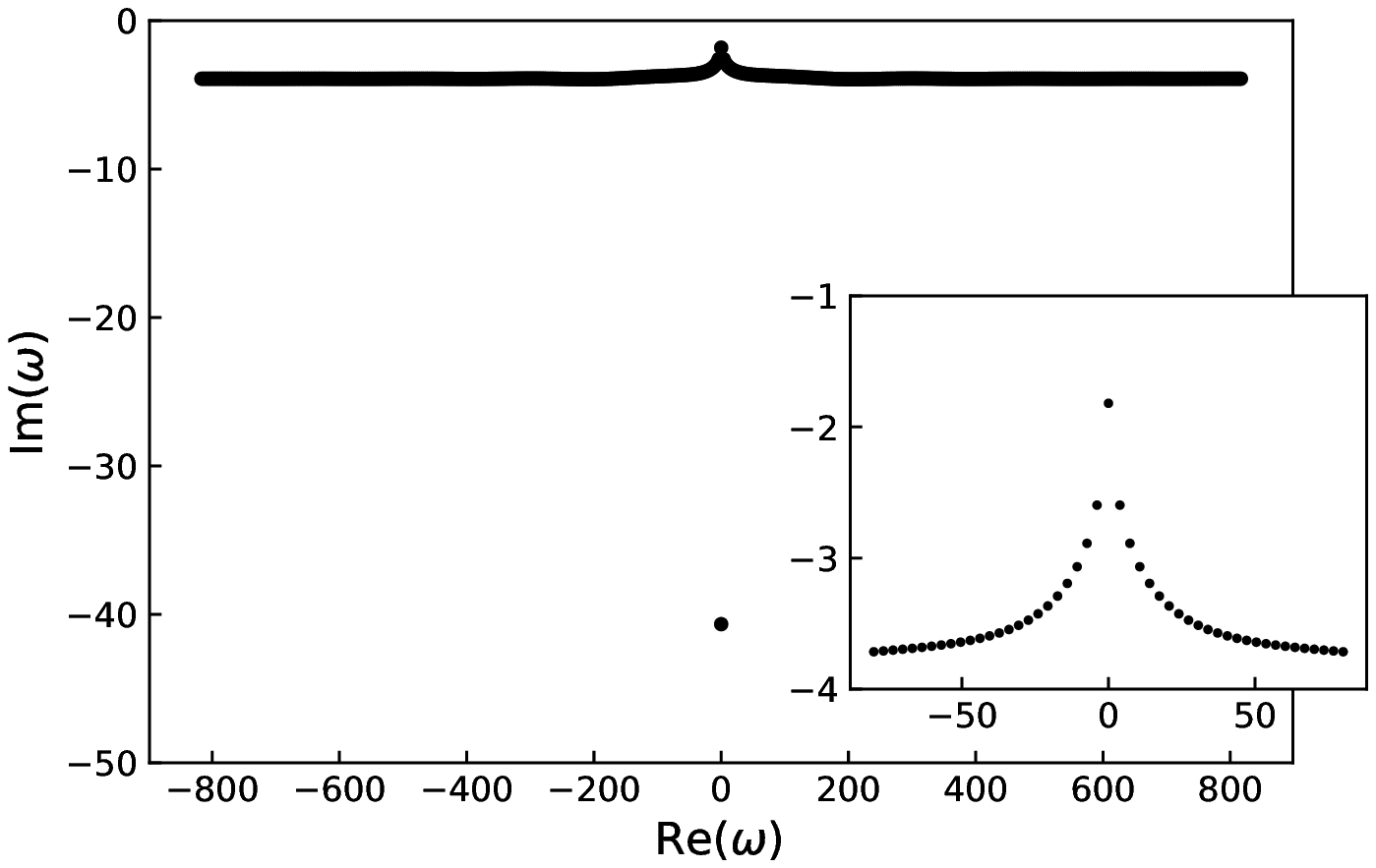}}
    \caption{Complex eigenspectra of the linear stability problem \eqref{stab_matrix}--\eqref{stab_bc2} at  different $\hat{Re}$ with $\lambda=0.5$ fixed.}
    \label{fig:eig1}
\end{figure}

\begin{figure}
    \centering
    \subfloat[$\lambda=0.1$]{\includegraphics[width=0.49\textwidth]{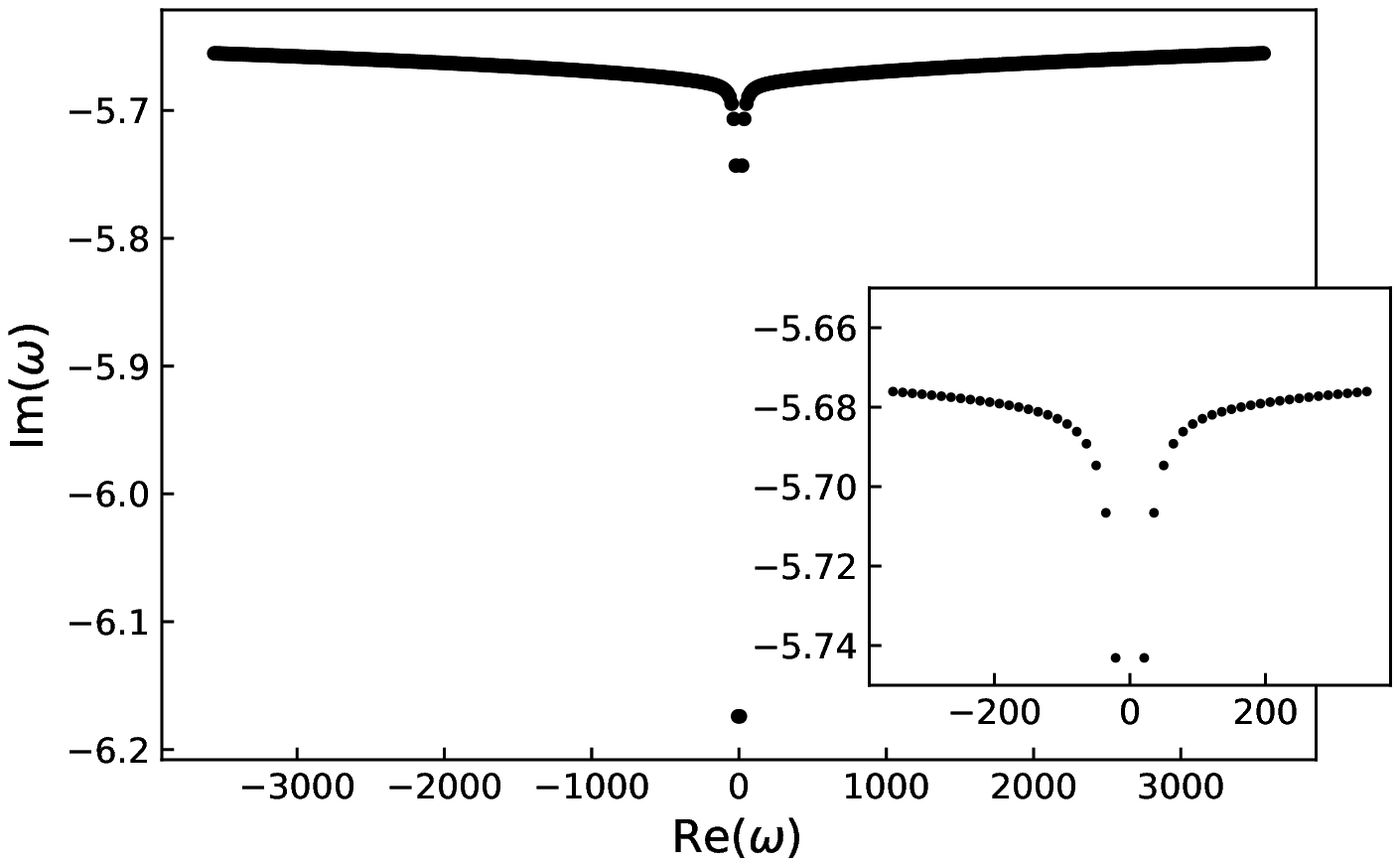}}
    \subfloat[$\lambda=1.0$]{\includegraphics[width=0.49\textwidth]{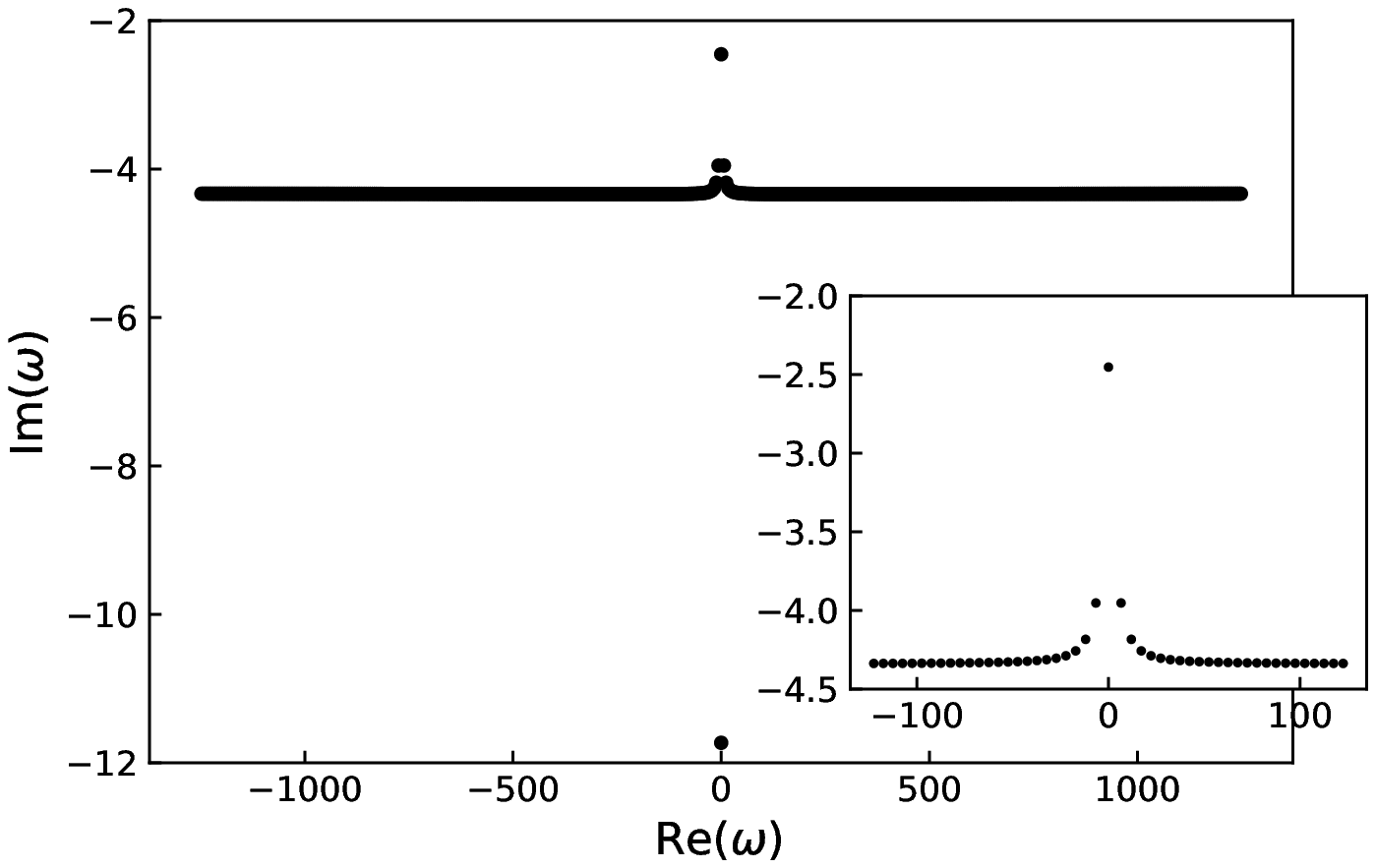}}
    \caption{Complex eigenspectra of the linear stability problem \eqref{stab_matrix}--\eqref{stab_bc2} for (a) $\lambda=0.1$ and (b) $\lambda=1.0$ with  $\hat{Re}=1$ fixed.}
    \label{fig:eig2}
\end{figure}

The eigenspectra for our genearalized eigenvalue problem are discrete. Since the left-hand-side matrix is real while the right-hand-matrix is purely imaginary, the resulting eigenspectra in $\mathbb{C}$ are symmetric about the imaginary axis. Multiple eigenvalue pairs, which are complex conjugates and thus have the same magnitude, are observed in our calculations (see Figs.~\ref{fig:eig1} and \ref{fig:eig2}). The eigenvalues are ordered with ascending magnitude and thus, the eigenvalue pairs  share the same position in the $\mathbb{C}$ plane. 

In the following discussion, different modes are referred to as the eigenfunctions corresponding to eigenvalues with different magnitudes $|\omega|$. For example, the first mode corresponds to the eigenvalue with the smallest magnitude, and the second mode has the eigenvalue with the second smallest magnitude, and so on. Furthermore, it is worth pointing out that, for our generalized eigenvalue problem \eqref{stab_matrix}, in principle, the eigenspectra should consist of an infinite number of discrete points, as the differential operators are infinite dimensional objects. However, since we numerically solve the problem by pseudospectral discretization, the resolution of the eigenspectra is determined by the number of Guass--Lobatto points. Therefore, considering the limits numerical linear algebra algorithms, the eigenspectra shown are the part for which the magnitudes of the eigenvalues are relatively small, whose computation is tractable using a finite number of grid points. The following results are calculated with $N=1000$ Gauss--Lobatto points for both eigenfunctions, $Q_1$ and $P_1$, with only the first 500 eigenvalues shown in Figs.~\ref{fig:eig1} and \ref{fig:eig2}. The accuracy of the calculations is assured by comparing the latter results to those with $N=800$ Gauss--Lobatto points for verification.

First, we investigate the eigenspectra by varying $\hat{Re}$ and fixing $\lambda=0.5$, as shown in Fig.~\ref{fig:eig1}. With this value of $\lambda$, appreciable deformation is observed in the base state (see Fig.~\ref{fig:P0H0}(b)). With the increase of $\hat{Re}$, ranging from $0.01$ to $3$, the imaginary parts of the majority of eigenvalues increase. However, no instabilities are observed as $\Imag(\omega)<0$ for all cases considered. Several modes with purely imaginary eigenvalues are found. Specifically, for $\hat{Re}=0.01$, there are 6 purely decaying modes, while only 2 such modes are observed for the other three cases. Among these modes, the one closest to the real axis is of interest because it represents the slowest decaying mode of the system. Table~\ref{tab:table-II} lists the largest imaginary part of all modes for the four values of $\hat{Re}$ considered. Interestingly, we do not observe a monotonic trend with the increase of $\hat{Re}$. Indeed, even without FSI, it is expected that a duct flow becomes more unstable as $\hat{Re}$ increases \cite{SH01}. 

Let us now take a look at the real part of the eigenvalues. For each case, the difference in the magnitudes of the real parts of two different modes is much larger than that of their imaginary parts, which is why the eigenspectra have a ``seagull" shape with a pair of relatively flat wings. The multiple eigenvalues with large-magnitude real parts evidence the existence of the highly-oscillatory eigenmodes, indicating the inherent stiffness of this FSI system. Comparing the cases of different $\hat{Re}$ in Fig.~\ref{fig:eig1}, the real parts of the eigenvalues display a decreasing trend with the increase of $\hat{Re}$.

Next, we keep $\hat{Re}=1$ fixed while varying the FSI parameter, $\lambda$. Note that our system is governed by two dimensionless groups, unlike classical hydrodynamics stability problems \cite{SH01}, which is the result of the coupled physics involved in two-way FSI. Still, as shown in Fig.~\ref{fig:eig2}, no instabilities are observed by varying $\lambda$, but there are some interesting differences with respect to varying $\hat{Re}$. For instance, in a less compliant system with $\lambda=0.1$, there are no purely decaying modes; all modes have non-zero real parts, meaning they are intrinsically oscillatory. It is also observed that $\Real(\omega)$ decreases as $\lambda$ increases.

\begin{figure}
    \centering
    \subfloat[]{\includegraphics[width=0.49\textwidth]{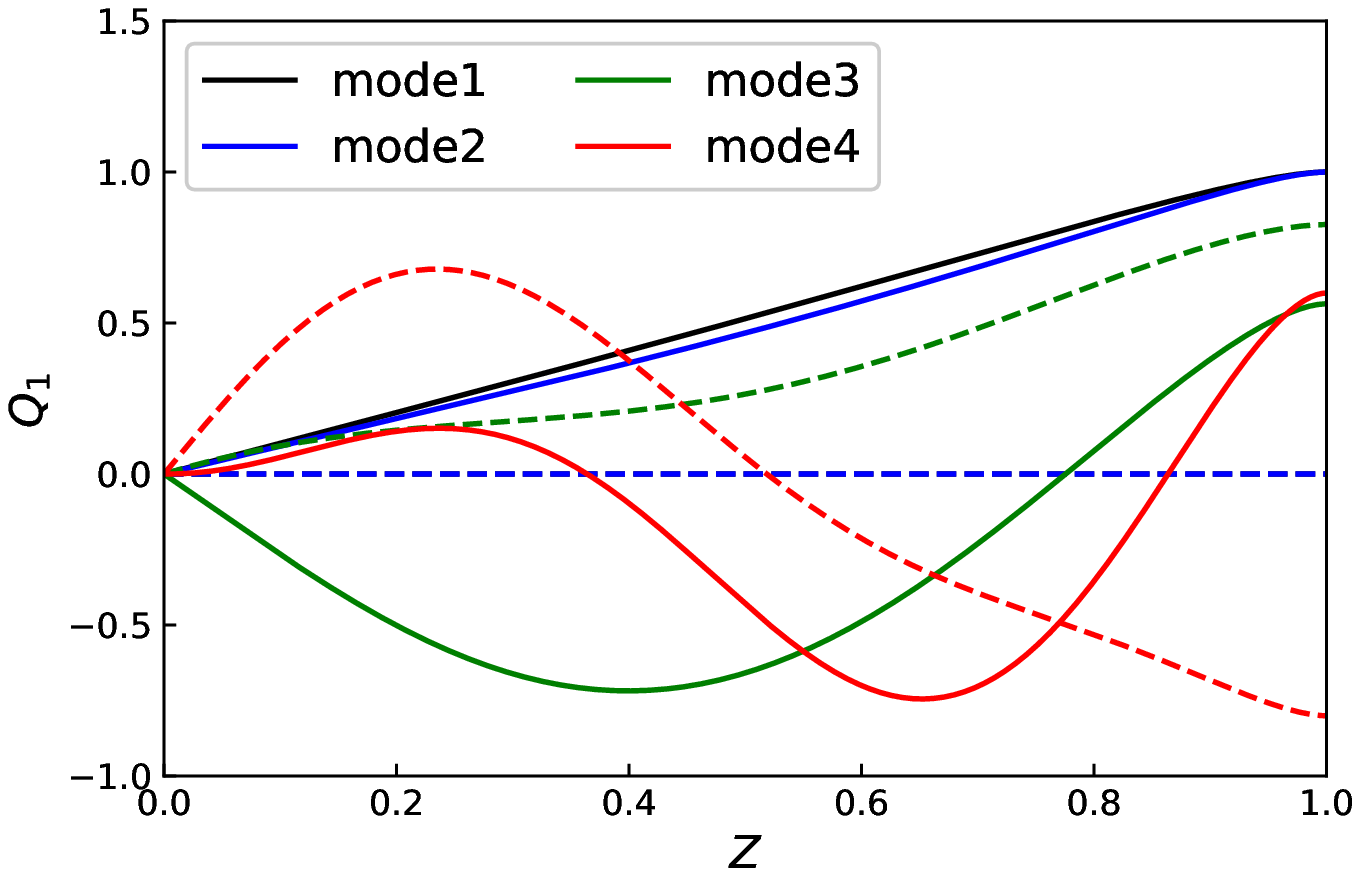}}
    \subfloat[]{\includegraphics[width=0.49\textwidth]{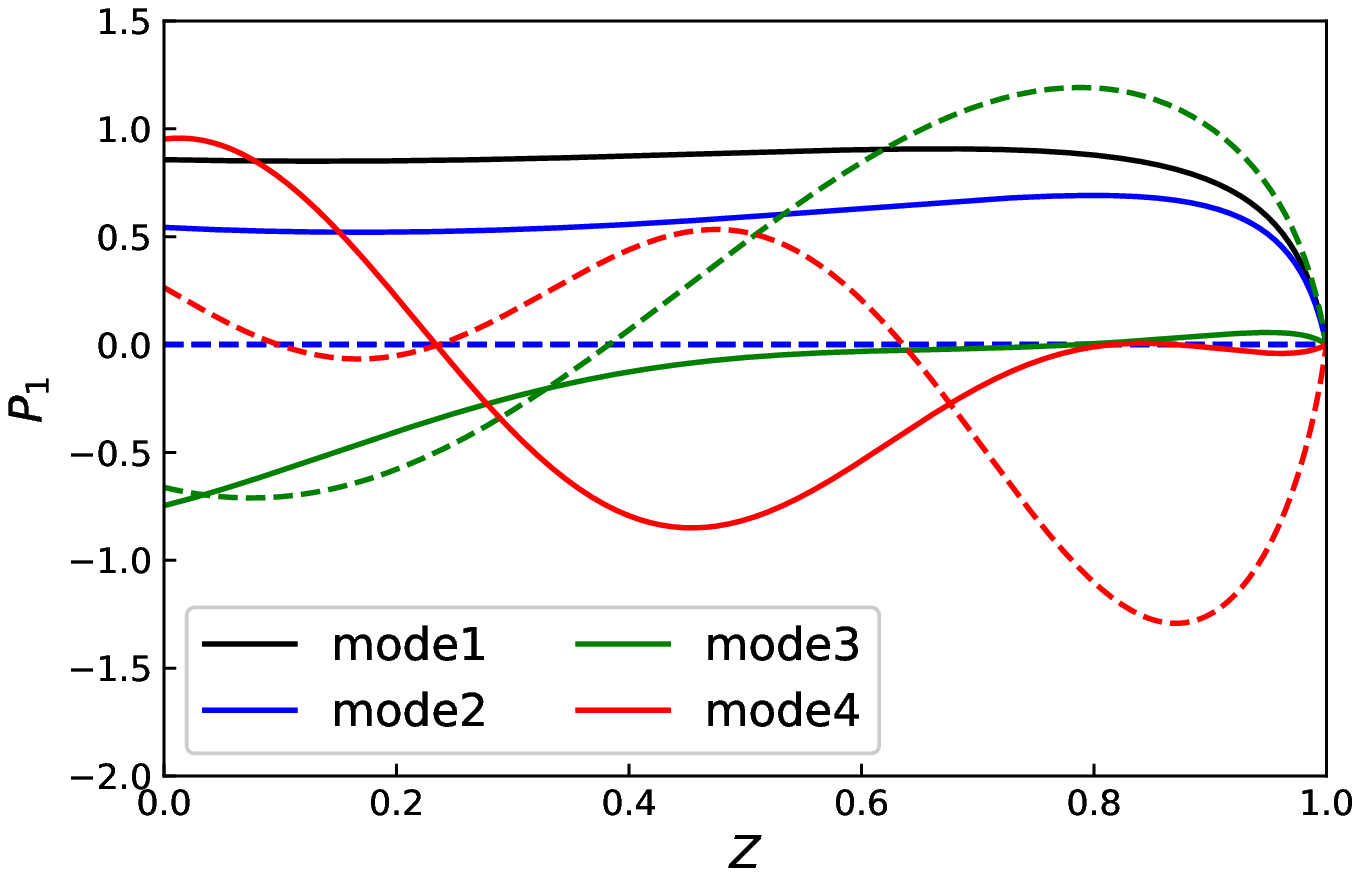}}
    \caption{Eigenfunctions: (a) $Q_1$ and (b) $P_1$ corresponding to the first four modes (ordered by $|\omega|$) for $\hat{Re}=1$ and $\lambda=0.5$. The solid curves represent the real part while the dashed curves represent the imaginary part of the eigenfunctions.}
    \label{fig:eigfunc}
\end{figure}

As for the eigenfunctions, in Fig.~\ref{fig:eigfunc} we show the first four modes for the case of $\hat{Re}=1$ and $\lambda=0.5$ as an example. The first two modes (labelled ``mode1" and ``mode2") correspond to two eigenvalues with $\Real(\omega)=0$ and $\Imag(\omega)<0$ from Fig.~\ref{fig:eig1}(c), for which the eigenfunctions are found to be real. In particular, $Q_1$ is monotonically increasing from the inlet to the outlet, while $P_1$ is relatively flat for most of the channel,  displaying a sharp decrease near the outlet. For the other two modes (labelled ``mode3" and ``mode4"), the corresponding eigenfunctions exhibit spatially-varying crests or troughs. The eigenfunctions of the fourth mode are ``wavier'' than the third mode. This observation is typical, and more humps would be observed in the higher modes, if we were to plot them.

\begin{table}
    \caption{\label{tab:table-II} The largest imaginary part of the eigenvalues corresponding to different $\hat{Re}$ with $\lambda=0.5$ fixed.}
    \begin{ruledtabular}
    \begin{tabular}{c c c c c}
    $\hat{Re}$  & 0.01 & 0.1 & 1 & 3\\
    \hline
    $\Imag(\omega)$& $-4.1459$ & $-4.1582$ & $-4.4004$ & $-1.8194$\\
    \end{tabular}
    \end{ruledtabular}
\end{table}

\section{Conclusion}

In this preliminary assessment of linear stability of the novel coupled flow and deformation solution in a thick-walled rectangular microchannel from Ref.~\cite{WC19}, we found that, within a range of the reduced Reynolds number, $\hat{Re}$, and the FSI parameter, $\lambda$, the inflated base state is linearly stable to infinitesimal flow-wise perturbations. With the Chebyshev pseudospectral method, we were able to resolve multiple highly oscillatory but stable eigenmodes, which indicates the stiffness of the FSI system. Unlike problems of classical hydrodynamic stability of duct flows \cite{SH01}, this FSI problem is governed by two dimensionless groups ($\hat{Re}$ and $\lambda$), and they both have a non-trivial influence on the eigenspectrum.

Importantly, unlike previous work \cite{VK13}, wherein the linear stability analysis was conducted locally on an approximately flat base state and the nonlinear pressure gradient was imported from computational fluid dynamics (CFD) simulations in a static but deformed geometry, the base state that we perturbed herein is non-flat, computed consistently from two-way coupled FSI theory \cite{WC19}. This base state, featuring a nonlinear pressure gradient, was derived from the closed-form 1D model consisting of Eqs.~\eqref{steady} and \eqref{steadybc}. Indeed, in Ref.~\cite{VK13}, the nonlinear pressure gradient was conjectured to be the most important factor in triggering instability. The velocity profile, which was chosen in Ref.~\cite{VK13} to be a quartic because it was closer to the output of  CFD simulations than the parabolic profile, was thought to be slightly less significant. In this respect, even though the velocity profile in our analysis is parabolic (to be consistent with the $\hat{Re}\to0$ solution \cite{WC19}), other profiles shapes (as function of $Y$) are allowed within the von K\'arm\'an--Polhausen approximation in Eq.~\eqref{Vz-Q}, as long as the boundary conditions at the top and bottom walls are satisfied.

Admittedly, our different formulation of the linear stability problem led us to a different conclusion from Ref.~\cite{VK13}, and we did not reproduce the instabilities observed therein. Nevertheless, the experiments \cite{VK13} are reproducible \cite{KB16} and the ultra-fast mixing phenomenon at low Reynolds number is striking. Therefore, the phenomenon of low-Reynolds-number FSI-induced instabilities remains relevant to understand from scratch (without ``infusing'' the linear stability calculation with CFD or experimental results) due to its potential relevance as new modality of mixing in microfluidics \cite{OW04}. One of the possible reasons that our reduced model did not predict an instability is that we fixed the upstream flow rate and set the outlet pressure to gauge. These boundary conditions might not perfectly match the experimental conditions in Ref.~\cite{VK13}. Another possibility may be that, the inertia of the elastic solid, which we have neglected, plays a role in triggering the instability.

In future work, we would like to address the effect of different boundary conditions on the linear (in)stability problem formulated herein. For example, we might consider fixing the total pressure drop $\Delta P$ across the length of the channel, leaving the inlet flow rate to be ``free.'' Another extension of the present theory can be accomplished by properly introducing the compliant wall's inertia (and unsteadiness) into the formulation. This extension requires updating the current solid mechanics model by properly justifying an independent time scale over which the elastic deformation varies.

\begin{acknowledgments}
This research was supported by the US National Science Foundation under grant No.\ CBET-1705637. I.C.C.\ is grateful to Prof.\ Michail Todorov for his invitation to present this research at the Twelfth Conference of the Euro-American Consortium for Promoting the Application of Mathematics in Technical and Natural Sciences (AMiTaNS'20). 
\end{acknowledgments}

\bibliography{mendeley_refs.bib}

\begin{thebibliography}{22}%
\makeatletter
\providecommand \@ifxundefined [1]{%
 \@ifx{#1\undefined}
}%
\providecommand \@ifnum [1]{%
 \ifnum #1\expandafter \@firstoftwo
 \else \expandafter \@secondoftwo
 \fi
}%
\providecommand \@ifx [1]{%
 \ifx #1\expandafter \@firstoftwo
 \else \expandafter \@secondoftwo
 \fi
}%
\providecommand \natexlab [1]{#1}%
\providecommand \enquote  [1]{``#1''}%
\providecommand \bibnamefont  [1]{#1}%
\providecommand \bibfnamefont [1]{#1}%
\providecommand \citenamefont [1]{#1}%
\providecommand \href@noop [0]{\@secondoftwo}%
\providecommand \href [0]{\begingroup \@sanitize@url \@href}%
\providecommand \@href[1]{\@@startlink{#1}\@@href}%
\providecommand \@@href[1]{\endgroup#1\@@endlink}%
\providecommand \@sanitize@url [0]{\catcode `\\12\catcode `\$12\catcode
  `\&12\catcode `\#12\catcode `\^12\catcode `\_12\catcode `\%12\relax}%
\providecommand \@@startlink[1]{}%
\providecommand \@@endlink[0]{}%
\providecommand \url  [0]{\begingroup\@sanitize@url \@url }%
\providecommand \@url [1]{\endgroup\@href {#1}{\urlprefix }}%
\providecommand \urlprefix  [0]{URL }%
\providecommand \Eprint [0]{\href }%
\providecommand \doibase [0]{http://dx.doi.org/}%
\providecommand \selectlanguage [0]{\@gobble}%
\providecommand \bibinfo  [0]{\@secondoftwo}%
\providecommand \bibfield  [0]{\@secondoftwo}%
\providecommand \translation [1]{[#1]}%
\providecommand \BibitemOpen [0]{}%
\providecommand \bibitemStop [0]{}%
\providecommand \bibitemNoStop [0]{.\EOS\space}%
\providecommand \EOS [0]{\spacefactor3000\relax}%
\providecommand \BibitemShut  [1]{\csname bibitem#1\endcsname}%
\let\auto@bib@innerbib\@empty
\bibitem [{\citenamefont {Pa{\"{i}}doussis}(2016)}]{P16}%
  \BibitemOpen
  \bibfield  {author} {\bibinfo {author} {\bibfnamefont {M.~P.}\ \bibnamefont
  {Pa{\"{i}}doussis}},\ }\href {\doibase 10.1016/C2011-0-08058-4} {\emph
  {\bibinfo {title} {{Fluid-Structure Interactions: Slender Structures and
  Axial Flow}}}},\ Vol.~\bibinfo {volume} {2}\ (\bibinfo  {publisher} {Academic
  Press},\ \bibinfo {address} {San Diego, CA},\ \bibinfo {year}
  {2016})\BibitemShut {NoStop}%
\bibitem [{\citenamefont {Bisplinghoff}\ \emph {et~al.}(1996)\citenamefont
  {Bisplinghoff}, \citenamefont {Ashley},\ and\ \citenamefont
  {Halfman}}]{BAH96}%
  \BibitemOpen
  \bibfield  {author} {\bibinfo {author} {\bibfnamefont {R.~L.}\ \bibnamefont
  {Bisplinghoff}}, \bibinfo {author} {\bibfnamefont {H.}~\bibnamefont
  {Ashley}}, \ and\ \bibinfo {author} {\bibfnamefont {R.~L.}\ \bibnamefont
  {Halfman}},\ }\href@noop {} {\emph {\bibinfo {title} {{Aeroelasticity}}}}\
  (\bibinfo  {publisher} {Dover Publications},\ \bibinfo {address} {Mineola,
  NY},\ \bibinfo {year} {1996})\BibitemShut {NoStop}%
\bibitem [{\citenamefont {Pedley}(1980)}]{P80}%
  \BibitemOpen
  \bibfield  {author} {\bibinfo {author} {\bibfnamefont {T.~J.}\ \bibnamefont
  {Pedley}},\ }\href {\doibase 10.1017/CBO9780511896996} {\emph {\bibinfo
  {title} {{The Fluid Mechanics of Large Blood Vessels}}}}\ (\bibinfo
  {publisher} {Cambridge University Press},\ \bibinfo {address} {Cambridge},\
  \bibinfo {year} {1980})\BibitemShut {NoStop}%
\bibitem [{\citenamefont {Chakraborty}\ \emph {et~al.}(2012)\citenamefont
  {Chakraborty}, \citenamefont {Prakash}, \citenamefont {Friend},\ and\
  \citenamefont {Yeo}}]{CPFY12}%
  \BibitemOpen
  \bibfield  {author} {\bibinfo {author} {\bibfnamefont {D.}~\bibnamefont
  {Chakraborty}}, \bibinfo {author} {\bibfnamefont {J.~R.}\ \bibnamefont
  {Prakash}}, \bibinfo {author} {\bibfnamefont {J.}~\bibnamefont {Friend}}, \
  and\ \bibinfo {author} {\bibfnamefont {L.}~\bibnamefont {Yeo}},\ }\href
  {\doibase 10.1063/1.4759493} {\bibfield  {journal} {\bibinfo  {journal}
  {Phys. Fluids}\ }\textbf {\bibinfo {volume} {24}},\ \bibinfo {pages} {102002}
  (\bibinfo {year} {2012})}\BibitemShut {NoStop}%
\bibitem [{\citenamefont {Duprat}\ and\ \citenamefont {Stone}(2016)}]{DS16}%
  \BibitemOpen
  \bibinfo {editor} {\bibfnamefont {C.}~\bibnamefont {Duprat}}\ and\ \bibinfo
  {editor} {\bibfnamefont {H.~A.}\ \bibnamefont {Stone}},\ eds.,\ \href
  {\doibase 10.1039/9781782628491} {\emph {\bibinfo {title} {{Fluid--Structure
  Interactions in Low-Reynolds-Number Flows}}}}\ (\bibinfo  {publisher} {The
  Royal Society of Chemistry},\ \bibinfo {address} {Cambridge, UK},\ \bibinfo
  {year} {2016})\BibitemShut {NoStop}%
\bibitem [{\citenamefont {Karan}\ \emph {et~al.}(2018)\citenamefont {Karan},
  \citenamefont {Chakraborty},\ and\ \citenamefont {Chakraborty}}]{KCC18}%
  \BibitemOpen
  \bibfield  {author} {\bibinfo {author} {\bibfnamefont {P.}~\bibnamefont
  {Karan}}, \bibinfo {author} {\bibfnamefont {J.}~\bibnamefont {Chakraborty}},
  \ and\ \bibinfo {author} {\bibfnamefont {S.}~\bibnamefont {Chakraborty}},\
  }\href {\doibase 10.1007/s41745-018-0073-5} {\bibfield  {journal} {\bibinfo
  {journal} {J. Indian Inst. Sci.}\ }\textbf {\bibinfo {volume} {98}},\
  \bibinfo {pages} {159} (\bibinfo {year} {2018})}\BibitemShut {NoStop}%
\bibitem [{\citenamefont {Fallahi}\ \emph {et~al.}(2019)\citenamefont
  {Fallahi}, \citenamefont {Zhang}, \citenamefont {Phan},\ and\ \citenamefont
  {Nguyen}}]{FZPN19}%
  \BibitemOpen
  \bibfield  {author} {\bibinfo {author} {\bibfnamefont {H.}~\bibnamefont
  {Fallahi}}, \bibinfo {author} {\bibfnamefont {J.}~\bibnamefont {Zhang}},
  \bibinfo {author} {\bibfnamefont {H.-P.}\ \bibnamefont {Phan}}, \ and\
  \bibinfo {author} {\bibfnamefont {N.-T.}\ \bibnamefont {Nguyen}},\ }\href
  {\doibase 10.3390/mi10120830} {\bibfield  {journal} {\bibinfo  {journal}
  {Micromachines}\ }\textbf {\bibinfo {volume} {10}},\ \bibinfo {pages} {830}
  (\bibinfo {year} {2019})}\BibitemShut {NoStop}%
\bibitem [{\citenamefont {Matia}\ \emph {et~al.}(2017)\citenamefont {Matia},
  \citenamefont {Elimelech},\ and\ \citenamefont {Gat}}]{MEG17}%
  \BibitemOpen
  \bibfield  {author} {\bibinfo {author} {\bibfnamefont {Y.}~\bibnamefont
  {Matia}}, \bibinfo {author} {\bibfnamefont {T.}~\bibnamefont {Elimelech}}, \
  and\ \bibinfo {author} {\bibfnamefont {A.~D.}\ \bibnamefont {Gat}},\ }\href
  {\doibase 10.1089/soro.2016.0048} {\bibfield  {journal} {\bibinfo  {journal}
  {Soft Robotics}\ }\textbf {\bibinfo {volume} {4}},\ \bibinfo {pages} {126}
  (\bibinfo {year} {2017})}\BibitemShut {NoStop}%
\bibitem [{\citenamefont {Verma}\ and\ \citenamefont {Kumaran}(2013)}]{VK13}%
  \BibitemOpen
  \bibfield  {author} {\bibinfo {author} {\bibfnamefont {M.~K.~S.}\
  \bibnamefont {Verma}}\ and\ \bibinfo {author} {\bibfnamefont
  {V.}~\bibnamefont {Kumaran}},\ }\href {\doibase 10.1017/jfm.2013.264}
  {\bibfield  {journal} {\bibinfo  {journal} {J. Fluid Mech.}\ }\textbf
  {\bibinfo {volume} {727}},\ \bibinfo {pages} {407} (\bibinfo {year}
  {2013})}\BibitemShut {NoStop}%
\bibitem [{\citenamefont {Kumaran}\ and\ \citenamefont {Bandaru}(2016)}]{KB16}%
  \BibitemOpen
  \bibfield  {author} {\bibinfo {author} {\bibfnamefont {V.}~\bibnamefont
  {Kumaran}}\ and\ \bibinfo {author} {\bibfnamefont {P.}~\bibnamefont
  {Bandaru}},\ }\href {\doibase 10.1016/j.ces.2016.04.001} {\bibfield
  {journal} {\bibinfo  {journal} {Chem. Eng. Sci.}\ }\textbf {\bibinfo {volume}
  {149}},\ \bibinfo {pages} {156} (\bibinfo {year} {2016})}\BibitemShut
  {NoStop}%
\bibitem [{\citenamefont {Wang}\ and\ \citenamefont {Christov}(2019)}]{WC19}%
  \BibitemOpen
  \bibfield  {author} {\bibinfo {author} {\bibfnamefont {X.}~\bibnamefont
  {Wang}}\ and\ \bibinfo {author} {\bibfnamefont {I.~C.}\ \bibnamefont
  {Christov}},\ }\href {\doibase 10.1098/rspa.2019.0513} {\bibfield  {journal}
  {\bibinfo  {journal} {Proc. R. Soc. A}\ }\textbf {\bibinfo {volume} {475}},\
  \bibinfo {pages} {20190513} (\bibinfo {year} {2019})}\BibitemShut {NoStop}%
\bibitem [{\citenamefont {Panda}\ \emph {et~al.}(2009)\citenamefont {Panda},
  \citenamefont {Yuet}, \citenamefont {Dendukuri}, \citenamefont {Hatton},\
  and\ \citenamefont {Doyle}}]{PYDHD09}%
  \BibitemOpen
  \bibfield  {author} {\bibinfo {author} {\bibfnamefont {P.}~\bibnamefont
  {Panda}}, \bibinfo {author} {\bibfnamefont {K.~P.}\ \bibnamefont {Yuet}},
  \bibinfo {author} {\bibfnamefont {D.}~\bibnamefont {Dendukuri}}, \bibinfo
  {author} {\bibfnamefont {T.~A.}\ \bibnamefont {Hatton}}, \ and\ \bibinfo
  {author} {\bibfnamefont {P.~S.}\ \bibnamefont {Doyle}},\ }\href {\doibase
  10.1088/1367-2630/11/11/115001} {\bibfield  {journal} {\bibinfo  {journal}
  {New J. Phys.}\ }\textbf {\bibinfo {volume} {11}},\ \bibinfo {pages} {115001}
  (\bibinfo {year} {2009})}\BibitemShut {NoStop}%
\bibitem [{\citenamefont {Mart{\'{i}}nez-Calvo}\ \emph
  {et~al.}(2020)\citenamefont {Mart{\'{i}}nez-Calvo}, \citenamefont {Sevilla},
  \citenamefont {Peng},\ and\ \citenamefont {Stone}}]{MCSPS19}%
  \BibitemOpen
  \bibfield  {author} {\bibinfo {author} {\bibfnamefont {A.}~\bibnamefont
  {Mart{\'{i}}nez-Calvo}}, \bibinfo {author} {\bibfnamefont {A.}~\bibnamefont
  {Sevilla}}, \bibinfo {author} {\bibfnamefont {G.~G.}\ \bibnamefont {Peng}}, \
  and\ \bibinfo {author} {\bibfnamefont {H.~A.}\ \bibnamefont {Stone}},\ }\href
  {\doibase 10.1017/jfm.2019.994} {\bibfield  {journal} {\bibinfo  {journal}
  {J. Fluid Mech.}\ }\textbf {\bibinfo {volume} {885}},\ \bibinfo {pages} {A25}
  (\bibinfo {year} {2020})}\BibitemShut {NoStop}%
\bibitem [{\citenamefont {Inamdar}\ \emph {et~al.}(2020)\citenamefont
  {Inamdar}, \citenamefont {Wang},\ and\ \citenamefont {Christov}}]{IWC20}%
  \BibitemOpen
  \bibfield  {author} {\bibinfo {author} {\bibfnamefont {T.~C.}\ \bibnamefont
  {Inamdar}}, \bibinfo {author} {\bibfnamefont {X.}~\bibnamefont {Wang}}, \
  and\ \bibinfo {author} {\bibfnamefont {I.~C.}\ \bibnamefont {Christov}},\
  }\href {\doibase 10.1103/PhysRevFluids.5.064101} {\bibfield  {journal}
  {\bibinfo  {journal} {Phys. Rev. Fluids}\ }\textbf {\bibinfo {volume} {5}},\
  \bibinfo {pages} {064101} (\bibinfo {year} {2020})}\BibitemShut {NoStop}%
\bibitem [{\citenamefont {Stewart}\ \emph {et~al.}(2009)\citenamefont
  {Stewart}, \citenamefont {Waters},\ and\ \citenamefont {Jensen}}]{SWJ09}%
  \BibitemOpen
  \bibfield  {author} {\bibinfo {author} {\bibfnamefont {P.~S.}\ \bibnamefont
  {Stewart}}, \bibinfo {author} {\bibfnamefont {S.~L.}\ \bibnamefont {Waters}},
  \ and\ \bibinfo {author} {\bibfnamefont {O.~E.}\ \bibnamefont {Jensen}},\
  }\href {\doibase 10.1016/j.euromechflu.2009.03.002} {\bibfield  {journal}
  {\bibinfo  {journal} {Eur. J. Mech. B/Fluids}\ }\textbf {\bibinfo {volume}
  {28}},\ \bibinfo {pages} {541} (\bibinfo {year} {2009})}\BibitemShut
  {NoStop}%
\bibitem [{\citenamefont {Panton}(2013)}]{panton}%
  \BibitemOpen
  \bibfield  {author} {\bibinfo {author} {\bibfnamefont {R.~L.}\ \bibnamefont
  {Panton}},\ }\href {\doibase 10.1002/9781118713075} {\emph {\bibinfo {title}
  {{Incompressible Flow}}}},\ \bibinfo {edition} {4th}\ ed.\ (\bibinfo
  {publisher} {John Wiley {\&} Sons},\ \bibinfo {address} {Hoboken, NJ},\
  \bibinfo {year} {2013})\BibitemShut {NoStop}%
\bibitem [{\citenamefont {Anand}\ and\ \citenamefont {Christov}(2020)}]{AC18b}%
  \BibitemOpen
  \bibfield  {author} {\bibinfo {author} {\bibfnamefont {V.}~\bibnamefont
  {Anand}}\ and\ \bibinfo {author} {\bibfnamefont {I.~C.}\ \bibnamefont
  {Christov}},\ }\href {\doibase 10.1002/zamm.201900309} {\bibfield  {journal}
  {\bibinfo  {journal} {Z. Angew. Math. Mech. (ZAMM)}\ } (\bibinfo {year}
  {2020}),\ 10.1002/zamm.201900309}\BibitemShut {NoStop}%
\bibitem [{\citenamefont {Virtanen}\ \emph {et~al.}(2020)\citenamefont
  {Virtanen}, \citenamefont {Gommers}, \citenamefont {Oliphant}, \citenamefont
  {Haberland}, \citenamefont {Reddy}, \citenamefont {Cournapeau}, \citenamefont
  {Burovski}, \citenamefont {Peterson}, \citenamefont {Weckesser},
  \citenamefont {Bright}, \citenamefont {van~der Walt}, \citenamefont {Brett},
  \citenamefont {Wilson}, \citenamefont {Millman}, \citenamefont {Mayorov},
  \citenamefont {Nelson}, \citenamefont {Jones}, \citenamefont {Kern},
  \citenamefont {Larson}, \citenamefont {Carey}, \citenamefont {Polat},
  \citenamefont {Feng}, \citenamefont {Moore}, \citenamefont {VanderPlas},
  \citenamefont {Laxalde}, \citenamefont {Perktold}, \citenamefont {Cimrman},
  \citenamefont {Henriksen}, \citenamefont {Quintero}, \citenamefont {Harris},
  \citenamefont {Archibald}, \citenamefont {Ribeiro}, \citenamefont
  {Pedregosa},\ and\ \citenamefont {van Mulbregt}}]{SciPy}%
  \BibitemOpen
  \bibfield  {author} {\bibinfo {author} {\bibfnamefont {P.}~\bibnamefont
  {Virtanen}}, \bibinfo {author} {\bibfnamefont {R.}~\bibnamefont {Gommers}},
  \bibinfo {author} {\bibfnamefont {T.~E.}\ \bibnamefont {Oliphant}}, \bibinfo
  {author} {\bibfnamefont {M.}~\bibnamefont {Haberland}}, \bibinfo {author}
  {\bibfnamefont {T.}~\bibnamefont {Reddy}}, \bibinfo {author} {\bibfnamefont
  {D.}~\bibnamefont {Cournapeau}}, \bibinfo {author} {\bibfnamefont
  {E.}~\bibnamefont {Burovski}}, \bibinfo {author} {\bibfnamefont
  {P.}~\bibnamefont {Peterson}}, \bibinfo {author} {\bibfnamefont
  {W.}~\bibnamefont {Weckesser}}, \bibinfo {author} {\bibfnamefont
  {J.}~\bibnamefont {Bright}}, \bibinfo {author} {\bibfnamefont {S.~J.}\
  \bibnamefont {van~der Walt}}, \bibinfo {author} {\bibfnamefont
  {M.}~\bibnamefont {Brett}}, \bibinfo {author} {\bibfnamefont
  {J.}~\bibnamefont {Wilson}}, \bibinfo {author} {\bibfnamefont {K.~J.}\
  \bibnamefont {Millman}}, \bibinfo {author} {\bibfnamefont {N.}~\bibnamefont
  {Mayorov}}, \bibinfo {author} {\bibfnamefont {A.~R.~J.}\ \bibnamefont
  {Nelson}}, \bibinfo {author} {\bibfnamefont {E.}~\bibnamefont {Jones}},
  \bibinfo {author} {\bibfnamefont {R.}~\bibnamefont {Kern}}, \bibinfo {author}
  {\bibfnamefont {E.}~\bibnamefont {Larson}}, \bibinfo {author} {\bibfnamefont
  {C.~J.}\ \bibnamefont {Carey}}, \bibinfo {author} {\bibfnamefont
  {I.}~\bibnamefont {Polat}}, \bibinfo {author} {\bibfnamefont
  {Y.}~\bibnamefont {Feng}}, \bibinfo {author} {\bibfnamefont {E.~W.}\
  \bibnamefont {Moore}}, \bibinfo {author} {\bibfnamefont {J.}~\bibnamefont
  {VanderPlas}}, \bibinfo {author} {\bibfnamefont {D.}~\bibnamefont {Laxalde}},
  \bibinfo {author} {\bibfnamefont {J.}~\bibnamefont {Perktold}}, \bibinfo
  {author} {\bibfnamefont {R.}~\bibnamefont {Cimrman}}, \bibinfo {author}
  {\bibfnamefont {I.}~\bibnamefont {Henriksen}}, \bibinfo {author}
  {\bibfnamefont {E.~A.}\ \bibnamefont {Quintero}}, \bibinfo {author}
  {\bibfnamefont {C.~R.}\ \bibnamefont {Harris}}, \bibinfo {author}
  {\bibfnamefont {A.~M.}\ \bibnamefont {Archibald}}, \bibinfo {author}
  {\bibfnamefont {A.~H.}\ \bibnamefont {Ribeiro}}, \bibinfo {author}
  {\bibfnamefont {F.}~\bibnamefont {Pedregosa}}, \ and\ \bibinfo {author}
  {\bibfnamefont {P.}~\bibnamefont {van Mulbregt}},\ }\href {\doibase
  10.1038/s41592-019-0686-2} {\bibfield  {journal} {\bibinfo  {journal} {Nature
  Methods}\ }\textbf {\bibinfo {volume} {17}},\ \bibinfo {pages} {261}
  (\bibinfo {year} {2020})}\BibitemShut {NoStop}%
\bibitem [{\citenamefont {Schmid}(2007)}]{Sch07}%
  \BibitemOpen
  \bibfield  {author} {\bibinfo {author} {\bibfnamefont {P.~J.}\ \bibnamefont
  {Schmid}},\ }\href {\doibase 10.1146/annurev.fluid.38.050304.092139}
  {\bibfield  {journal} {\bibinfo  {journal} {Annu. Rev. Fluid Mech.}\ }\textbf
  {\bibinfo {volume} {39}},\ \bibinfo {pages} {129} (\bibinfo {year}
  {2007})}\BibitemShut {NoStop}%
\bibitem [{\citenamefont {Schmid}\ and\ \citenamefont
  {Henningson}(2001)}]{SH01}%
  \BibitemOpen
  \bibfield  {author} {\bibinfo {author} {\bibfnamefont {P.~J.}\ \bibnamefont
  {Schmid}}\ and\ \bibinfo {author} {\bibfnamefont {D.~S.}\ \bibnamefont
  {Henningson}},\ }\href {\doibase 10.1007/978-1-4613-0185-1} {\emph {\bibinfo
  {title} {{Stability and Transition in Shear Flows}}}},\ \bibinfo {series}
  {Applied Mathematical Sciences}, Vol.\ \bibinfo {volume} {142}\ (\bibinfo
  {publisher} {Springer},\ \bibinfo {address} {New York, NY},\ \bibinfo {year}
  {2001})\BibitemShut {NoStop}%
\bibitem [{\citenamefont {Boyd}(2000)}]{Boyd00}%
  \BibitemOpen
  \bibfield  {author} {\bibinfo {author} {\bibfnamefont {J.~P.}\ \bibnamefont
  {Boyd}},\ }\href@noop {} {\emph {\bibinfo {title} {{Chebyshev and Fourier
  Spectral Methods}}}},\ \bibinfo {edition} {2nd}\ ed.\ (\bibinfo  {publisher}
  {Dover Publications},\ \bibinfo {address} {Mineola, NY},\ \bibinfo {year}
  {2000})\BibitemShut {NoStop}%
\bibitem [{\citenamefont {Ottino}\ and\ \citenamefont {Wiggins}(2004)}]{OW04}%
  \BibitemOpen
  \bibfield  {author} {\bibinfo {author} {\bibfnamefont {J.~M.}\ \bibnamefont
  {Ottino}}\ and\ \bibinfo {author} {\bibfnamefont {S.}~\bibnamefont
  {Wiggins}},\ }\href {\doibase 10.1098/rsta.2003.1355} {\bibfield  {journal}
  {\bibinfo  {journal} {Phil. Trans. R. Soc. A}\ }\textbf {\bibinfo {volume}
  {362}},\ \bibinfo {pages} {923} (\bibinfo {year} {2004})}\BibitemShut
  {NoStop}%
\end{thebibliography}%

\end{document}